\documentclass[twocolumn, twoside]{IEEEtran}

\usepackage{graphicx}
\usepackage{amsmath}
\usepackage{amssymb}
\usepackage{amsthm}

\newtheorem{theorem}{Theorem}
\newtheorem{lemma}{Lemma}
\newtheorem{corollary}{Corollary}

\newtheorem{example}{Example}

\begin{document}

\title{Frequency-Hopping Sequence Sets With Low Average and Maximum
Hamming Correlation}

\author{Jin-Ho Chung and Kyeongcheol Yang
\thanks{This work was supported by the National Research
Foundation of Korea (NRF) grant funded by the Korea government
(MEST) (No. 2011-0017396).}
\thanks{J.-H. Chung and K. Yang are with the Dept.
   of Electrical Engineering,
   Pohang University of Science and Technology (POSTECH),
   Pohang, Kyungbuk 790-784, Korea (e-mail: {jinho, kcyang}@postech.ac.kr).}
  }
{\markboth{Submitted to IEEE Transactions on Information Theory,
July 27, 2011}{Chung and Yang: Frequency-hopping sequence sets
with low average and maximum Hamming correlation} \maketitle

\begin{abstract}

In frequency-hopping multiple-access (FHMA) systems, the average
Hamming correlation (AHC) among frequency-hopping sequences (FHSs)
as well as the maximum Hamming correlation (MHC) is an important
performance measure. Therefore, it is a challenging problem to
design FHS sets with good AHC and MHC properties for application.
In this paper, we analyze the AHC properties of an FHS set, and
present new constructions for FHS sets with optimal AHC. We first
calculate the AHC of some known FHS sets with optimal MHC, and
check their optimalities. We then prove that any uniformly
distributed FHS set has optimal AHC. We also present two
constructions of FHS sets with optimal AHC based on cyclotomy.
Finally, we show that if an FHS set is obtained from another FHS
set with optimal AHC by an interleaving, it has optimal AHC.
\end{abstract}


\section{Introduction}

 In multiple-access communication systems, the receiver is
confronted with the interference caused by undesired signals when
it attempts to demodulate one of the signals sent from several
transmitters. For frequency-hopping multiple-access (FHMA)
systems, such a multiple-access interference (MAI) arise mainly
from the hits of frequencies assigned to users in each time slot.
It is possible to reduce the MAI in multiple-access systems by
employing frequency-hopping sequence (FHS) sets with low Hamming
correlation. There are two measures on the Hamming correlation of
an FHS set used in FHMA systems. The average Hamming correlation
(AHC) among FHSs measures its average performance, while the
maximum Hamming correlation (MHC) represents its worst-case
performance. Therefore, AHC as well as MHC is an important
performance measure for an FHS set.

In general, it is desirable that an FHS set should have a large
set size and a low AHC or MHC value, when its length and the
number of available frequencies are fixed. There are several known
constructions \cite{Lempel}-\cite{Chung11} for FHS sets having
optimal MHC with respect to the Peng-Fan bound \cite{Peng0}. On
the other hand, only a few constructions for FHS sets with optimal
AHC have been known because AHC has been
 recently considered \cite{Peng1}. Peng {\em et al}. in
\cite{Peng1} established a bound on the AHC of an FHS set and
presented some FHS sets with optimal AHC, which are based on cubic
polynomials. By using the theory of generalized cyclotomy
\cite{Whiteman}, Liu {\em et al}. also constructed FHS sets with
optimal AHC \cite{Liu}. Unfortunately, some previously known FHS
sets with optimal AHC do not have good MHC properties. Therefore,
it is a challenging problem to design FHS sets with optimal AHC
and low MHC.

In this paper, we deal with FHS sets having optimal AHC and low
MHC. We check the relation between optimal MHC and AHC by
calculating the AHC values of some known optimal FHS sets with
respect to the Peng-Fan bound. We also show that any `uniformly
distributed' FHS set has optimal AHC, and present some examples
with low MHC. We then construct some FHS sets with optimal AHC and
low MHC based on cyclotomy, which have lengths $p$ or $p^n-1$ for
a prime $p$ and a positive integer $n$. Finally, we analyze the
optimality of FHS sets constructed by interleaving techniques, and
give some new interleaved FHS sets with optimal AHC and low MHC.

The outline of the paper is as follows. In Section II, some
preliminaries on AHC and MHC are presented. We prove that a
uniformly distributed FHS set has optimal AHC and give some
examples in Section III. In Section IV, we present new
constructions for FHS sets with optimal AHC based on cyclotomy.
The AHC properties of FHS sets constructed by interleaving
techniques are analyzed in Section V. Finally, we give some
concluding remarks in Section VI.

\medskip

\section{Average Hamming Correlation of FHS Sets}

 Throughout the paper, we denote by $\lceil
x \rceil$ the smallest integer greater than or equal to $x$. We
also denote by $\langle x \rangle_y$ the least nonnegative residue
of $x$ modulo $y$ for an integer $x$ and a positive integer $y$.
For an integer $m$, we denote by $\mathbb{Z}_m$ the set of
integers modulo $m$.

\vspace{-0.1in}

\subsection{Maximum Hamming Correlation}

Let $\mathcal F=\{f_0, f_1, \ldots, f_{M-1}\}$ be a set of
available frequencies. A sequence $X=\{X(t)\}_{t=0}^{N-1}$ is
called an FHS of length $N$ over ${\cal F}$ if $X(t)\in {\cal F}$
for all $0\le t\le N-1$. For two FHSs $X$ and $Y$ of length $N$
over ${\cal F}$, the {\em periodic Hamming correlation} between
$X$ and $Y$ is defined as
\[
    H_{X,Y}(\tau)=\sum_{t=0}^{N-1}h[X(t),Y(\langle t+\tau \rangle_N)],
    ~~~~0\le\tau\le N-1
\]
where
\[
     h[x,y]=
     \left\{ \begin{array}{ll}
         1, &\text{if ~$x=y$}\,\\
         0, &\text{otherwise.}
     \end{array} \right.
\]
If $X=Y$, $H_{X,Y}(\tau)$ is called the {\em Hamming
autocorrelation} of $X$, denoted by $H_X(\tau)$.  The {\em maximum
out-of-phase Hamming autocorrelation} of $X$ is defined as
\[
    H(X)=\max_{1\le\tau\le N-1}\{H_X(\tau)\}.
\]

\noindent A well-known bound on it, called the {\em
Lempel-Greenberger bound} \cite{Lempel}, is given in the following
lemma.

\setcounter{lemma}{0}
\begin{lemma} [\cite{Lempel}]
    For any FHS $X$ of length $N$ over $\mathcal F$ with
    $|\mathcal F|=M$,
    \begin{equation}
        H(X) \ge \left\lceil \frac{(N-b)(N+b-M)}{M(N-1)} \right\rceil
    \label{LG}
    \end{equation}
    where $b=\langle N\rangle_M$.
    \label{Bound1}
\end{lemma}

 Let ${\cal U}$ be
an $(N,M,L)$-FHS set, that is, an FHS set consisting of $L$ FHSs
of length $N$ over ${\cal F}$. For any two distinct FHSs $X$ and
$Y$ in ${\cal U}$, let
    \[
        H(X,Y)=\max_{0\le\tau\le N-1}\{H_{X,Y}(\tau)\}.
    \]

\noindent The {\em maximum out-of-phase Hamming autocorrelation}
$H_{\rm a}({\cal U})$ and the {\em maximum Hamming
crosscorrelation} $H_{\rm c}({\cal U})$ of ${\cal U}$ are defined
as
\begin{eqnarray*}
    &&H_{\rm a}({\cal U})=\max_{X\in\,{\cal U}}\{H(X)\},\\
    &&H_{\rm c}({\cal U})=\max_{X,Y\in\,{\cal U},\, X\ne Y}\{H(X,Y)\},
\end{eqnarray*}
respectively. The {\em maximum Hamming correlation} of $\cal U$ is
also defined as
\[
    H(\mathcal U)=\max\{H_{\rm a}(\mathcal U),H_{\rm c}(\mathcal
    U)\}.
\]

Peng and Fan established some bounds on the maximum out-of-phase
Hamming autocorrelation and the maximum Hamming crosscorrelation
of an FHS set in terms of frequency set size, length, and the
number of FHSs \cite{Peng0}.

\setcounter{lemma}{1}
\begin{lemma} [\cite{Peng0}]
    Let ${\cal U}$ be an $(N,M,L)$-FHS set. Then
    \begin{eqnarray}
        && M(N-1)H_{\rm a}({\cal U})+NM(L-1)H_{\rm c}({\cal U})
           ~~~~~~~~~~~~  \nonumber \\
        && ~~~~~~~~  \ge  N(NL-M). \label{PF1}
    \end{eqnarray}
    \label{PFbound}
\end{lemma}

An FHS set ${\cal U}$ is said to have {\em optimal MHC} if
$(H_{\rm a}({\cal U}), H_{\rm c}({\cal U}))=(\lambda_{\rm a},
\lambda_{\rm c})$, where the integer pair $(\lambda_{\rm
a},\lambda_{\rm c})$ satisfies (\ref{PF1}), but $(\lambda_{\rm
a}-\delta, \lambda_{\rm c}-\delta)$ does not satisfy (\ref{PF1})
for any positive integer $\delta$ \cite{Chung11}. The set ${\cal
U}$ is said to have {\em near-optimal} MHC if $(H_{\rm a}({\cal
U}), H_{\rm c}({\cal U}))=(\lambda_{\rm a}+1, \lambda_{\rm c}+1)$.

\subsection{Average Hamming Correlation}

Let ${\cal U}$ be an $(N,M,L)$-FHS set. For our convenience, let
$S_{\rm a}({\cal U})$ be the sum of all out-of-phase Hamming
autocorrelation values in ${\cal U}$, that is,
\[
    S_{\rm a}({\cal U})=\sum_{X\in\,{\cal U}}\,
    \sum_{\tau=1}^{N-1}
    H_X(\tau).
\]
Similarly, let ${\cal S}_{\rm c}({\cal U})$ be the sum of all
Hamming crosscorrelation values in ${\cal U}$, given by
\[
    S_{\rm c}({\cal U})
    =\sum_{X,\,Y\in\,{\cal U},\, X\ne Y}
    \sum_{\tau=0}^{N-1}
    H_{X,Y}(\tau).
\]
Then, the average Hamming autocorrelation and crosscorrelation of
${\cal U}$ are defined by
\[
    A_{\rm a}({\cal U})
    =\frac{S_{\rm a}({\cal U})}{L(N-1)}
\]
and
\[
    A_{\rm c}({\cal U})
    =\frac{S_{\rm c}({\cal U})}
                    {L(L-1)N},
\]
respectively. Peng {\em et al}. established a bound on the AHC of
an FHS set \cite{Peng1}.

\setcounter{lemma}{2}
\begin{lemma}[\cite{Peng1}]
    Let ${\cal U}$ be an $(N,M,L)$-FHS set. Then
    \begin{equation}
        \frac{A_{\rm a}({\cal U})}{N(L-1)}
        +\frac{A_{\rm c}({\cal U})}{N-1}
        \ge
        \frac{NL-M}{M(N-1)(L-1)}.
        \label{PNT_bound}
    \end{equation}
    \label{PNT_Lemma}
\end{lemma}

An FHS set ${\cal U}$ will be said to have {\em optimal AHC} if
the pair $(A_{\rm a}({\cal U}),A_{\rm c}({\cal U}))$ satisfies
(\ref{PNT_bound}) with equality. Note that an optimal pair
satisfying the bound in (\ref{PNT_bound}) may consist of rational
numbers, while every optimal pair satisfying the Peng-Fan bound is
an integer pair. Moreover, AHC is not directly related to MHC from
the viewpoints of their definitions. In fact, an optimal FHS set
with respect to the Peng-Fan bound is not necessarily optimal with
respect to the AHC bound in (\ref{PNT_bound}). Therefore, it is
interesting to investigate the AHC properties of known FHS sets
with optimal MHC. The AHC values of some known FHS sets with
optimal MHC are calculated and summarized in Table I.

\medskip

\noindent\textbf{Remark}: In Theorem 2 of \cite{Peng1}, Peng {\em
et al}. mentioned without a detailed proof that any FHS set with
optimal MHC has optimal AHC, assuming that $(A_{\rm a}({\cal
U}),A_{\rm c}({\cal U}))$ is an integer pair. However, their
argument is not true in general since the assumption is not always
valid, as discussed above.

\bigskip

\section{Average Hamming Correlation of Uniformly Distributed
FHS Sets}

Balancedness is one of the major randomness measures for
deterministically generated sequences, since it is closely related
to unpredictability \cite{Golomb}. Hence, it is very important to
design balanced sequences and analyze their correlation properties
for their application to communication systems. In this section,
we investigate the AHC properties of a special class of FHS sets,
called `uniformly distributed' FHS sets, and present some examples
of them which also have low MHC.

Given an FHS $X=\{X(t)\}_{t=0}^{N-1}$ over ${\cal F}$, let
\[
    N_X(a)=|\{t\,:\,X(t)=a,~0\le t\le N-1\}|
\]
for $a\in {\cal F}$.  When $|N_X(a)-N_X(b)|\le 1$ for any $a,b\in
{\cal F}$, we call $X$ a {\em balanced} FHS. In particular, $X$
will be referred to as a {\em perfectly balanced} FHS if
$|N_X(a)-N_X(b)|=0$ for any $a,b\in {\cal F}$. Note that $N$
should be a multiple of the size of ${\cal F}$ if $X$ is perfectly
balanced. An FHS set consisting of (perfectly) balanced FHSs is
called a {\em (perfectly) balanced} FHS set.

The following well-known identity gives a relationship between the
distribution of frequencies and the sum of Hamming correlation
values between two FHSs. It is very useful in deriving some bounds
on the MHC or AHC of an FHS set, including the Peng-Fan bound
\cite{Peng0}.

\setcounter{lemma}{3}
\begin{lemma}
    Let $X=\{X(t)\}_{t=0}^{N-1}$ and  $Y=\{Y(t)\}_{t=0}^{N-1}$ be
    two FHSs over ${\cal F}$. Then
    \[
        \sum_{\tau=0}^{N-1}H_{X,Y}(\tau)=
        \sum_{a\in{\cal F}}N_X(a)N_Y(a).
    \]
    In particular,
    \[
        \sum_{\tau=0}^{N-1}H_{X}(\tau)=
        \sum_{a\in{\cal F}}N_X(a)^2.
    \]
    \label{identity}
\end{lemma}

Given an $(N,M,L)$-FHS set ${\cal X}=\{X_i~|~0\le i\le L-1\}$ over
${\cal F}$, define
\[
    N_{\cal X}(a)=\sum_{i=0}^{L-1}N_{X_i}(a)
\]
for $a\in {\cal F}$. The FHS set ${\cal X}$ is called a {\em
uniformly distributed} FHS set if $|N_{\cal X}(a)-N_{\cal
X}(b)|=0$ for any $a,b\in {\cal F}$. In this case, it is required
that $M\,|\,NL$. Clearly, an FHS set is uniformly distributed  if
it is perfectly balanced. The following lemma gives a relation
between the sums of Hamming correlation values of ${\cal X}$ and
the numbers $N_{\cal X}(a)$, $a\in {\cal F}$.

\setcounter{lemma}{4}
\begin{lemma}
    Let ${\cal X}$ be an $(N,M,L)$-FHS set over ${\cal F}$. Then
    \[
        S_{\rm a}({\cal X})+S_{\rm c}({\cal X})
        \,=\, \sum_{a\in \, {\cal F}}N_{\cal X}(a)
              \cdot \left(N_{\cal X}(a)-1\right).
    \]
    \label{identity_set}
\end{lemma}

{\em Proof}. Note that
    \begin{eqnarray*}
        S_{\rm a}({\cal X})+S_{\rm c}({\cal X})&=&
        \sum_{0\le\, i,\,j\le\, L-1}\, \sum_{\tau=0}^{N-1}\,
        H_{X_i,X_j}(\tau) \\
        && ~~-\, \sum_{0\,\le i\le \,L-1}\,H_{X_i}(0).
    \end{eqnarray*}
By applying Lemma \ref{identity} and the fact that
$H_{X_i}(0)$$=\sum_{a\in\,{\cal F}}N_{X_i}(a)$, we obtain
    \begin{eqnarray*}
        && S_{\rm a}({\cal X})+S_{\rm c}({\cal X}) \\
        && ~~ = \sum_{0\le i,\,j\le L-1}\, \sum_{a\in\,{\cal F}}\,
            N_{X_i}(a)N_{X_j}(a)\\
        && ~~~~~~ -\,
            \sum_{0\le
            i\le L-1}\,\sum_{a\in\,{\cal F}}\,N_{X_i}(a)\\
        && ~~ = \sum_{a\in\,{\cal F}} \left(
            \sum_{0\le \, i\le\, L-1}\,
            N_{X_i}(a)
            \sum_{0\le \, j\le\, L-1}\,
            N_{X_j}(a)\right. \\
        && ~~~~~~~~~~~~~~ -
            \left. \sum_{0\le i\le L-1}\, N_{X_i}(a)
            \right)\\
        && ~~ = \sum_{a\in\,{\cal F}} \left(\sum_{0\le \, i\le\, L-1}\,
            N_{X_i}(a)\right) \\
        && ~~~~~~~~~~~~ \cdot \left(\sum_{0\le \, j\le\, L-1}\,
            N_{X_i}(a)-1\right) \\
        && ~~ = \sum_{a\in \, {\cal F}}\, N_{\cal X}(a)
              \cdot \left(N_{\cal X}(a)-1\right)
    \end{eqnarray*}
where the last equality directly comes from the definition of
$N_{\cal X}(a)$. \hfill$\Box$

\bigskip

By Lemma \ref{identity_set}, it is possible to prove the
optimality of a uniformly distributed FHS set.

\setcounter{theorem}{5}
\begin{theorem}
    Let ${\cal X}$ be a uniformly
    distributed $(N,M,L)$-FHS set. Then ${\cal X}$ has optimal AHC.
    \label{Optimlity_uni}
\end{theorem}

{\em Proof}. Note that $M|\,NL$ and $N_{\cal X}(a)=\frac{NL}{M}$
for all $a\in{\cal F}$ since ${\cal X}$ is a uniformly distributed
FHS set. By Lemma \ref{identity_set}, we have
    \begin{eqnarray*}
        S_{\rm a}({\cal X})+ S_{\rm c}({\cal X})
        &=& M\cdot \frac{NL}{M}\cdot\left(\frac{NL}{M}-1\right)\\
        &=& \frac{NL(NL-M)}{M}.
    \end{eqnarray*}
Therefore, the left-hand side (LHS) and the right-hand side (RHS)
of (\ref{PNT_bound}) are given by
    \begin{eqnarray*}
        \text{LHS}&=&
        \frac{S_{\rm a}({\cal X})+S_{\rm c}({\cal X})}
        {NL(N-1)(L-1)}\\
        &=&\frac{NL-M}{M(N-1)(L-1)}
    \end{eqnarray*}
and
    \[
        \text{RHS}=\frac{NL-M}{M(N-1)(L-1)},
    \]
respectively. \hfill$\Box$

\smallskip

\setcounter{corollary}{6}
\begin{corollary}
    Let ${\cal X}$ be a perfectly balanced
    $(N,M,L)$-FHS set. Then ${\cal X}$ has optimal AHC.
    \label{Optimlity}
\end{corollary}

\smallskip

Theorem \ref{Optimlity_uni} and Corollary \ref{Optimlity} tell us
that any uniformly distributed or perfectly balanced FHS set is
optimal with respect to the bound in (\ref{PNT_bound}). However,
such an FHS set is also required to have good MHC properties if it
is applicable to FHMA systems. In the following, we will give
three examples of uniformly distributed or perfectly balanced FHS
sets with low MHC.

\smallskip

\setcounter{example}{7}
\begin{example}
    Let $p$ be an odd prime. For $0\le t\le p^2-1$, let
    $t=t_0p+t_1$, where $0\le t_0, t_1 \le p-1$.
    Let ${\cal X}_1$ be the $(p^2,p,p)$-FHS set
    defined as ${\cal X}_1=\{X_i~|~0\le i\le p-1\}$
    where $X_i=\{X_i(t)\}_{t=0}^{p^2-1}$ is the FHS over
    $p\,\mathbb{Z}_p$, given by
    \[
        X_i(t)=\langle p\,t_0t_1 + pi \rangle_{p^2}.
    \]
    It was proved by Kumar \cite{Kumar} that ${\cal X}_1$ is optimal
    with respect to the Peng-Fan bound. Note that ${\cal X}_1$
    is a uniformly distributed FHS set, since $N_{{\cal X}_1}(a)=p^2$
    for any $a\in p\,\mathbb{Z}_p$. Therefore,
    ${\cal X}_1$ has optimal AHC by Theorem \ref{Optimlity_uni}.
\end{example}

\smallskip

\setcounter{example}{8}
\begin{example}
    Let $k$ and $N$ be two positive integers such that $N\ge 3$
    and $2\le k<N$. Assume that $N=Ld+r$, where $L$ is a positive
    integer, $0\le r<d$, and
    $1\le d<\frac{N}{2}$. For an integer $0\le i\le L-1$,
    let $X_i=\{X_i(t)\}_{t=0}^{kN-1}$ be the
    FHS over $\mathbb{Z}_k\times \mathbb{Z}_N$ defined as
    \[
        X_i(kt_1+t_0)=\left\{\begin{array}{l}
                        (t_0,t_1+id), \\
                        ~~~~~~\text{ if }t_0=0, 1,\ldots,
                        \left\lfloor\frac{k}{2}\right\rfloor \\
                        (t_0,t_1+(L-1-i)d),\\
                        ~~~~~~ \text{ if } t_0=
                        \left\lfloor\frac{k}{2}\right\rfloor+1,
                        \ldots, k-1
                    \end{array}\right.
    \]
    where $0\le t_0 \le k-1$ and $0\le t_1\le N-1$.
    It was shown by Chung  et al. \cite{Chung1} that
    the $(kN,kN,L)$-FHS set ${\cal X}_2=\{X_i~|~0\le i\le
    L-1\}$ has zero Hamming autocorrelation for any $0<|\tau |<d-1$, and
    zero Hamming crosscorrelation for any $0\le |\tau | <d-1$, that is,
    it is a no-hit-zone FHS set. In particular, it is optimal with
    respect to the bound given in \cite{Ye1}. Note that
    ${\cal X}_2$ is a perfectly balanced FHS set, since
    $N_{X_i}((a,b))=1$ for any $0\le i\le L-1$ and any $(a,b)\in
    \mathbb{Z}_k\times \mathbb{Z}_N$. Therefore,
    ${\cal X}_2$ has optimal AHC by Corollary \ref{Optimlity}.
\end{example}

\smallskip

\setcounter{example}{9}
\begin{example}
    Let $p$ be an odd prime. For $0\le t\le p^2-p-1$, let $t_0=\langle
    t \rangle_{p-1}$ and $t_1=\langle t \rangle_{p}$. Let ${\cal
    X}_3$ be the $(p^2-p,p,p)$-FHS set
    defined as ${\cal X}_3=\{X_i~|~0\le i\le p-1\}$
    where $X_i=\{X_i(t)\}_{t=0}^{p^2-p-1}$ and
    \[
        X_i(t)=\langle (t_0+1)\cdot t_1 + i \rangle_p.
    \]
    It was proved in \cite{Chung2} that ${\cal X}_3$ is optimal
    with respect to the Peng-Fan bound. Note that  ${\cal X}_3$
    is a perfectly balanced FHS set, since $N_{X_i}(a)=p-1$ for
    any $0\le i\le p-1$ and any $a\in\mathbb{Z}_p$. Therefore,
    ${\cal X}_3$ has optimal AHC by Corollary \ref{Optimlity}.
\end{example}

\bigskip

\section{FHS Sets With Optimal Average Hamming Correlation Based
on Cyclotomy}

 Let $\mathbb F_{q}$ be the finite field of $q=p^n$
elements and $\mathbb F_{q}^*$ the set of nonzero elements in
$\mathbb F_{q}$ where $p$ is a prime and $n$ is a positive
integer. For some positive integers $M$ and $f$, let $q=Mf+1$. For
a primitive element $\alpha$ of $\mathbb F_{q}$, $\mathbb F_{q}^*$
is decomposed into $M$ disjoint subsets
\begin{equation*}
    C_r=\left\{\alpha^{Ml+r} ~|~0 \le l \le f-1 \right\},
    ~r=0,1,\ldots,M-1
\end{equation*}
which are called the {\em cyclotomic classes} of $\mathbb F_{q}$
of order $M$. For two integers $i$ and $j$ in $\mathbb{Z}_M$, the
number defined by
\begin{equation*}
(i,j)_M :=|(C_i+1) \cap C_j\,|
\end{equation*}
is called a {\em cyclotomic number} of $\mathbb F_{q}$ of order
$M$ \cite{Storer}.

The result in the following lemma was first proven by Sze {\em et
al.} in \cite{Sze} and was rediscovered by Chu and Colburn in
\cite{Chu}.

\setcounter{lemma}{10}
\begin{lemma}[\cite{Sze}]
     Let $q=Mf+1$ be a prime power. Then we have
     \[
              \sum_{i=0}^{M-1}(i+j,i)_M=
              \left\{\begin{array}{ll}
                f-1, & \text{ if }\,j\equiv 0~\text{\rm mod }M\\
                f, & \text{ if }\,j\ne 0 ~\text{\rm mod }M.
              \end{array}\right.
     \]
     \label{cyclotomy2}
\end{lemma}

In \cite{Chu}, Chu and Colburn gave optimal FHSs over
$\mathbb{Z}_M$ or $\mathbb{Z}_M \cup \{\infty\}$ with respect to
the Lempel-Greenberger bound as well as an optimal FHS set over
$\mathbb{Z}_M \cup \{\infty\}$ with respect to the Peng-Fan bound.
Although the FHS set has optimal MHC, it does not have optimal AHC
with respect to the bound in (\ref{PNT_bound}). In order to get an
FHS set with optimal AHC and near-optimal MHC, we modify their
construction as follows:

\medskip

\noindent \textbf{Construction A}: Let $p=Mf+1$ be an odd prime
for some positive integers $M$ and $f$. For $0\le i\le M-1$,
define the FHS $X_i=\{X_i(t)\}_{t=0}^{p-1}$ as
    \[
        X_i(0)=i
    \]
and
    \[
        X_i(t)=\left\langle r+i \right\rangle_M
        ~~\text{ if }t\in C_r
    \]
where $C_r$ is a cyclotomic class of $\mathbb{F}_p$ of order $M$.
Let ${\cal X}_4$ be the $(p,M,M)$-FHS set defined by
    \[
        {\cal X}_4=\{X_i~|~0\le i\le M-1\}.
    \]

By using the theory of cyclotomy \cite{Storer}, it is possible to
calculate the AHC and MHC values of ${\cal X}_4$ and check its
optimality.

\setcounter{theorem}{11}
\begin{theorem}
    The set ${\cal X}_4$ in Construction A has optimal AHC with respect
    to the bound in {\rm (\ref{PNT_bound})}. Moreover,
    it is near-optimal with respect to the Peng-Fan bound.
    \label{PRseq}
\end{theorem}

{\em Proof}. Clearly, the set ${\cal X}_4$ is a uniformly
distributed FHS set, and so it has optimal AHC by Theorem
\ref{Optimlity_uni}. Let $H_{i,j}(\tau)$ be the Hamming
correlation between $X_i$ and $X_j$. It is obvious that
$H_{i,j}(0)=p$ if $i=j$, and $H_{i,j}(0)=0$, otherwise. For $1\le
\tau \le p-1$, we have
    \begin{eqnarray*}
        H_{i,j}(\tau)&=& \sum_{t\in\,\mathbb{Z}_p\setminus\{-\tau,\,0\} }
        h[X_i(t),X_j(t+\tau)] \\
        && ~~ +I\left( \tau\in C_{i-j}\right)+
        I\left( -\tau\in C_{j-i}\right)\\
        &=& \sum_{r=0}^{M-1}(r,r+(j-i))_M \\
        && ~~  + \left|\{\tau\}\cap C_{i-j}\right|
        + \left|\{-\tau\}\cap C_{j-i}\right|.
    \end{eqnarray*}
Hence,
    \[
        H_{\rm a}({\cal X}_4)= f+1
    \]
and
    \[
        H_{\rm c}({\cal X}_4)= f+2
    \]
by Lemma \ref{cyclotomy2}. It is easily checked that ${\cal X}_4$
has near-optimal MHC. \hfill$\Box$

\bigskip

\noindent\textbf{Remark}:  The AHC values of ${\cal X}_4$ may be
easily computed by applying Lemma \ref{identity} to $S_{\rm
a}({\cal X}_4)$ and $S_{\rm c}({\cal X}_4)$. The average Hamming
autocorrelation of ${\cal X}_4$ is calculated as follows:
\begin{eqnarray*}
    A_{\rm a}({\cal X}_4)
    &=& \frac{1}{M(p-1)}\,\sum_{X\in\,{\cal X}_4}
        \,\sum_{\tau=1}^{p-1}\,H_X(\tau)\\
    &=& \frac{1}{M^2f}\sum_{X\in\,{\cal X}_4}
        \,\left[\,\sum_{a\in{\cal F}}\,N_X(a)^2-N_X(0)\right]\\
    &=&\frac{M\left((f+1)^2+(M-1)f^2-(Mf+1)\right)}{M^2f}\\
    &=& f-1+\frac{2}{M}\,.
\end{eqnarray*}
Similarly, the average Hamming crosscorrelation is given by
\begin{eqnarray*}
    A_{\rm c}({\cal X}_4)&=&
    \frac{1}{M(M-1)p}\,\sum_{X,\,Y\in\,{\cal X}_4,\,X\ne Y}
        \,\sum_{\tau=0}^{p-1}\,H_{X,\,Y}(\tau)\\
    &=& \frac{1}{M(M-1)p}\,\sum_{X,\,Y\in\,{\cal X}_4,\,X\ne Y}
       \,\sum_{a\in{\cal F}}\,N_X(a)N_Y(a)\\
    &=&\frac{M(M-1)(2f(f+1)+(M-2)f^2)}{M(M-1)(Mf+1)}\\
    &=& \frac{Mf^2+2f}{Mf+1}.
\end{eqnarray*}
It is easily checked that the pair $(A_{\rm a}({\cal X}_4), A_{\rm
c}({\cal X}_4))$ satisfies the bound in (\ref{PNT_bound}) with
equality.

\medskip

\noindent\textbf{Remark}:  Each FHS in ${\cal X}_4$ is optimal
with respect to the Lempel-Greenberger bound \cite{Lempel}.

\bigskip

In \cite{Ding}, Ding and Yin gave FHS sets of length $q-1$ over
$\mathbb{Z}_M$ or $\mathbb{Z}_M\cup \{\infty \}$ based on the
discrete logarithm in $\mathbb{F}_q$ for a prime power $q$.
 Han and Yang observed in \cite{Han}
that these FHS sets are closely related to Sidel'nikov sequences
and some of the results in \cite{Sidelnikov} and \cite{Ding} are
not correct, and made corrections to them. The FHS set over
$\mathbb{Z}_M$ by Ding and Yin \cite{Ding} can be equivalently
represented as follows:

\medskip

\noindent\textbf{Construction B}: For a prime power $q=p^n$ such
that there exist two integers $M$ and $f$ such that $q=Mf+1$, let
$C_r$, $0\le r\le M-1$ be the cyclotomic class of order $M$ of
$\mathbb{F}_q$. Let ${\cal X}_5$ be the $(q-1,M,M)$-FHS set over
$\mathbb{Z}_M$ given by
    \[
        {\cal X}_5=\left\{X_i~|~X_i=\{X_i(t)\}_{t=0}^{q-2},
        ~~0\le i\le M-1\right\}
    \]
where
    \[
        X_i(t)=\left\{\begin{array}{ll}
                    \langle r+i \rangle_M, &
                    \text{ if }\alpha^t+1\in C_r \\
             i,& \text{ if }\alpha^t+1=0.
                \end{array}\right.
    \]

\medskip

\setcounter{theorem}{12}
\begin{theorem}
    The set ${\cal X}_5$ in Construction B has optimal AHC
    and $H({\cal X}_5)\le f+2$.
    In particular, ${\cal X}_5$ has near-optimal MHC if
    $(2l,l)_M=0$ for all
    $l\in\mathbb{Z}_M$.
    \label{Sidel}
\end{theorem}

{\em Proof}. Clearly, the set ${\cal X}_5$ is a perfectly balanced
FHS set, and so it has optimal AHC by Theorem \ref{Optimlity_uni}.
Let $H_{i,j}(\tau)$ be the Hamming correlation between $X_i$ and
$X_j$ in ${\cal X}_5$. It is obvious that $H_{i,j}(0)=q-1$ if
$i=j$, and $H_{i,j}(0)=0$, otherwise. For $1\le \tau\le q-2$,
$H_{i,j}(\tau)$ is given by
    \begin{eqnarray*}
        H_{i,j}(\tau)&=&\sum_{r=0}^{M-1}(j+\tau,i)_M
         +I\left(1-\alpha^{\tau}\in C_{i-j}\right) \\
        && ~~ +I\left(-\alpha^{-\tau}\left(1-
        \alpha^{\tau}\in C_{j-i}\right)\right).
    \end{eqnarray*}
After some calculation, it is checked that ${\cal X}_5$ is a
near-optimal FHS set with respect to the Peng-Fan bound when
$\left|C_{l}\cap (C_{2l}+1)\right|=0$, that is, $(2l,l)_M=0$ for
all $l\in\mathbb{Z}_M$. \hfill$\Box$

\begin{table*}[tbp]
\begin{center}
    {\footnotesize\caption{AHC of FHS Sets With Low MHC (Here,
    $p,p_1,\ldots,p_r$ are odd primes. UD and PB
    denote `uniformly distributed' and `perfectly balanced',
    respectively.)}}
    \begin{tabular}{|c||c|c|c|c|c|c|c|c|}
        \hline
        Reference &
        $N$
        & $|{\cal F}|$ & $\begin{array}{c}
                \text{Distribution} \\
                \text{(FHS;}\\
                \text{FHS set)}
            \end{array}$ & $L$ &
        MHC &
        $A_{\rm a}({\cal X})$
        & $A_{\rm c}({\cal X})$ & AHC  \\
        \hline
        \hline
        \cite{Lempel}   & $p^m-1$ & $p$ &
         $\begin{array}{c}
                \text{balanced;} \\
                \text{UD}
            \end{array}$ &$p^{m-1}$ & optimal & $p^{m-1}-1$
        & $\frac{p^{m-1}(p^m-2)}{p^m-1}$ & optimal \\
        \hline
        \cite{Kumar} & $p^2$ & $p$ & $\begin{array}{c}
                \text{unbalanced;} \\
                \text{UD}
            \end{array}$ & $p$  &
        optimal & $p$ & $\frac{p^2-1}{p}$ & optimal \\
        \hline
        \cite{Chu} & $p=Mf+1$ & $M+1$ &
        $\begin{array}{c}
                \text{unbalanced;} \\
                \text{not UD}
            \end{array}$
        &$M$  & optimal & $\frac{p-M+1}{M}$
        & $\frac{Mf^2+2f+1}{p}$ & not optimal \\
        \hline
        \cite{Ding}, \cite{Han} & $p^m-1=Mf+1$ & $M+1$ &
        $\begin{array}{c}
                \text{unbalanced;} \\
                \text{not UD}
            \end{array}$
        &$M$  & optimal & $\frac{Mf^2-2f+2}{Mf-1}$
        & $\frac{Mf^2-2f-1}{Mf}$
        & not optimal \\
        \hline
        \cite{Chung1} &
        $\begin{array}{c}
                kN, \\
                2\le k<N
            \end{array}$ & $kN$ &
        $\begin{array}{c}
                \text{PB;}\\
                \text{PB}
            \end{array}$ & $N/k$ & not optimal
        & $0$ & $k$ & optimal \\
        \hline
        \cite{Chung2} & $p^2-p$ & $p$ &
        $\begin{array}{c}
                \text{PB;}\\
                \text{PB}
            \end{array}$ & $p$ & optimal
        & $\frac{p(p-1)(p-2)}{p^2-p-1}$ & $p$  & optimal \\
        \hline
        Theorem \ref{PRseq} & $p=Mf+1$ & $M$ &
        $\begin{array}{c}
                \text{balanced;} \\
                \text{UD}
            \end{array}$& $M$ & near-optimal
        & $\frac{p-M+1}{M}$ & $\frac{Mf^2+2f}{p}$ & optimal \\
                \hline
        Theorem \ref{Sidel} & $p^m-1=Mf$ & $M$ &
        $\begin{array}{c}
                \text{PB;}\\
                \text{PB}
            \end{array}$ & $M$ & near-optimal
        & $\frac{(f-1)(p^m-1)}{p^m-2}$ & $f$ & optimal \\
        \hline
        Theorem \ref{kN} &
        $\begin{array}{c}
                kp_1^{e_1}\cdots p_r^{e_r}, \\
                p_1<\cdots <p_r, \\
                e_1,\ldots, e_r\ge 1
            \end{array}$ &
        $p_1^{e_1}\cdots p_r^{e_r}$ &
        $\begin{array}{c}
                \text{PB;}\\
                \text{PB}
            \end{array}$ & $\frac{p_1-1}{k}$ & optimal &
        $\frac{k(k-1)N}{kN-1}$ & $k$ & optimal \\
        \hline
    \end{tabular}
\end{center}
\end{table*}

\bigskip
\section{Average Hamming Correlation of FHS Sets Based on
Interleaving Techniques}

Interleaving techniques are used to construct a sequence of length
$kN$ from $k$ sequences of length $N$, which are not necessarily
distinct for some positive integers $k$ and $N$ \cite{Gong0}. They
have been widely employed in the construction of sequences with
low correlation \cite{Gong0}, \cite{Gong1}, \cite{Zhou}. In
particular, Chung {\em et al}. \cite{Chung0} firstly applied the
interleaving techniques to the design of FHSs, and presented
several FHS sets with optimal MHC. A similar approach was given in
\cite{Chung1} to construct no-hit-zone FHS sets. However, these
previous works dealt only with the MHC of FHS sets constructed by
interleaving techniques. In this section, we will focus on the AHC
of such FHS sets. First of all, we will show that  the sum of
Hamming correlation values and the optimality on the AHC of an FHS
set are preserved under any interleaving in the following theorem.

\smallskip

\setcounter{theorem}{13}
\begin{theorem}
    Let ${\cal X}\triangleq \{X_i~|~0\le i\le L-1\}$ be an
    $(N,M,L)$-FHS set over ${\cal F}$ and
    ${\cal Y}\triangleq \{Y_j~|~0\le j\le L'-1\}$ an
    $(N',M,L')$-FHS set obtained by interleaving ${\cal X}$ such
    that $NL=N'L'$ and the FHS $~Y_j\triangleq\{Y_j(s)\}_{s=0}^{N'-1}$,
    $0\le j\le L'-1$, is defined as
    \[
        Y_j(s)=X_i(t)
    \]
    for some $0\le i\le L-1$ and $0\le t\le N-1$.
    Assume that $i_1=i_2$ and $t_1=t_2$ if
    and only if $j_1=j_2$ and $s_1=s_2$, when $Y_{j_1}(s_1)=X_{i_1}(t_1)$ and
    $Y_{j_2}(s_2)=X_{i_2}(t_2)$. Then, we have
    \begin{equation}
        S_{\rm a}({\cal X})+S_{\rm c}({\cal X})
        =S_{\rm a}({\cal Y})+S_{\rm c}({\cal Y}).
        \label{preserve}
    \end{equation}
    Furthermore, ${\cal X}$ has optimal AHC if and only if
    ${\cal Y}$ has optimal AHC.
    \label{interleaving_gen}
\end{theorem}

{\em Proof}. By the assumption on ${\cal X}$ and ${\cal Y}$, we
have $N_{\cal X}(a)=N_{\cal Y}(a)$ for all $a\in{\cal F}$. Hence,
$S_{\rm a}({\cal X})+S_{\rm c}({\cal X})=S_{\rm a}({\cal
Y})+S_{\rm c}({\cal Y})$ by Lemma \ref{identity_set}. Let ${\rm
LHS}_{\cal X}$ (resp. ${\rm LHS}_{\cal Y}$) be the left-hand side
of (\ref{PNT_bound}) for the FHS set ${\cal X}$ (resp. ${\cal
Y}$), and ${\rm RHS}_{\cal X}$ (resp. ${\rm RHS}_{\cal Y}$) the
right-hand side of (\ref{PNT_bound}) for ${\cal X}$ (resp. ${\cal
Y}$). By (\ref{preserve}) and the assumption that $NL=N'L'$, we
have
\begin{eqnarray*}
    {\rm LHS}_{\cal Y}&=&\frac{S_{\rm a}({\cal Y})+
    S_{\rm c}({\cal Y})}{N'L'(N'-1)(L'-1)}\\
    &=&\frac{S_{\rm a}({\cal X})+
    S_{\rm c}({\cal X})}{NL(N'-1)(L'-1)}\\
    &=& {\rm RHS}_{\cal X}\cdot
        \frac{(N-1)(L-1)}{(N'-1)(L'-1)}
\end{eqnarray*}
and
\begin{eqnarray*}
    {\rm RHS}_{\cal Y}&=&\frac{N'L'-M}{M(N'-1)(L'-1)}\\
    &=& {\rm RHS}_{\cal X}\cdot
        \frac{(N-1)(L-1)}{(N'-1)(L'-1)}.
\end{eqnarray*}
Therefore, ${\cal X}$ has optimal AHC if and only if ${\cal Y}$
has optimal AHC. \hfill$\Box$

\bigskip

\noindent\textbf{Remark}: The assumption on the indices $i,j,s$,
and $t$ in Theorem \ref{interleaving_gen} guarantees that $X_i(t)$
appears exactly once in ${\cal Y}$ through interleaving for any
$0\le i\le L-1$ and $0\le t\le N-1$.

\bigskip

The most common interleaving  is applied to an FHS set in the
following construction.

\medskip

\noindent\textbf{Construction C}: Let ${\cal
X}=\{X_0,\ldots,X_{L-1}\}$ be an $(N,M,L)$-FHS set. Let $k$ be a
positive integer such that $2\le k\le L$ and $k\,|\,L$. For $0\le
i\le L/k-1$, the FHS $Y_i=\{Y_i(t)\}_{t=0}^{kN-1}$ is defined as
    \[
        Y_i(kt_1+t_0)=X_{ki+t_0}(t_1)
    \]
where $0\le t_0\le k-1$ and $0\le t_1\le N-1$. The set ${\cal
Y}\triangleq \{Y_i\,|\,0\le i\le L/k-1\}$ is an $(kN,M,L/k)$-FHS
set.

\medskip

\setcounter{corollary}{14}
\begin{corollary}
    Let ${\cal X}$ be an FHS set with optimal AHC. Then, ${\cal
    Y}$ in Construction C is an FHS set with optimal AHC.
    \label{Thm_interleaved}
\end{corollary}

\medskip

Several FHS sets with low MHC were constructed by interleaving
techniques  in \cite{Chung0}. Theorem \ref{interleaving_gen} tells
us that interleaving techniques may also be a good design tool for
optimal AHC. We give a construction example by applying
Construction C to ${\cal X}_4$ in Construction A as follows.

\smallskip

\setcounter{corollary}{15}
\begin{corollary}
    Let $p=Mf+1$ be an odd prime for a positive integer $M$ and an
    odd integer $f$. Let ${\cal X}_4=\{X_i~|~0\le i\le M-1\}$ be
    the $(p,M,M)$-FHS set given in Construction A. For $0\le i\le
    M/2-1$, let $Y_i=\{Y_i(t)\}_{t=0}^{2p-1}$ be the FHS defined as
    \[
        Y_i(2t_1+t_0)=X_{2i+t_0}(t_1)
    \]
    where $0\le t_0 \le 1$, $0\le t_1 \le p-1$.
    Then ${\cal Y}_4\triangleq\{Y_i~|~0\le i\le M/2-1\}$ is a
    $(2p,M,M/2)$-FHS set with optimal AHC.
\end{corollary}

\medskip

By extending Construction B1 in \cite{Chung0}, it is also possible
to obtain an FHS set with optimal AHC and MHC.

\setcounter{theorem}{16}
\begin{theorem}
    Let $N=p_1^{e_1}\cdots p_r^{e_r}$ where $r\ge 1$, $e_i\ge 1$
    for all $1\le i\le r$, and $p_1<\cdots <p_r$ are odd primes.
    Define ${\cal X}\triangleq \{X_i~|~0\le i\le
    p_1-2\}$ as the $(N,N,p_1-1)$-FHS set over $\mathbb{Z}_N$,
    where
    \[
        X_i(t)=\left\langle (i+1)t \right\rangle_N.
    \]
    for $0\le t\le N-1$. For a positive divisor $k$ of $p_1-1$,
    let ${\cal Y}\triangleq\{Y_i~|~0\le i\le (p_1-1)/k-1\}$ be defined as
    \[
        Y_i(kt_1+t_0)=X_{ki+t_0}(t_1)
    \]
    where $0\le t_0\le k-1$ and $0\le t_1\le N-1$.
    Then, ${\cal Y}$ is a $\left(kN,N,(p_1-1)/k \right)$-FHS set
with optimal AHC and MHC.
    \label{kN}
\end{theorem}

{\em Proof}. Note that ${\cal X}$ is perfectly balanced, since
$N_{X_i}(a)=1$ for all $a\in \mathbb{Z}_N$ and all $0\le i\le
p_1-2$. This implies that ${\cal X}$ has optimal AHC by Corollary
\ref{Optimlity}. Hence, ${\cal Y}$ also has optimal AHC by
Corollary \ref{Thm_interleaved}. The MHC of ${\cal Y}$ can be
derived from the results of \cite{Cao} and \cite{Chung0}. In
\cite{Cao}, it was shown that
    \[
        H_{X_i,X_j}(\tau)=
        \left\{\begin{array}{ll}
                        N, & \text{if }~i=j\text{ and }\tau=0\\
                        0, & \text{if }~i=j\text{ and }\tau\ne 0\\
                        1, & \text{if }~i\ne j.
        \end{array}\right.
    \]
By extending the Proof of Theorem 15 in \cite{Chung0}, we obtain
 $H_{\rm a}({\cal Y})=H_{\rm c}({\cal Y})=k$. Therefore,
 ${\cal Y}$ is optimal with respect to the
Peng-Fan bound.  \hfill$\Box$

\smallskip

\section{Conclusion}

 Some known FHS sets with optimal MHC were classified by
their AHC properties. It was shown that any uniformly distributed
FHS set has optimal AHC. Two FHS constructions with optimal AHC
and near-optimal MHC were presented by using the theory of
cyclotomy. The optimality of FHS sets obtained by interleaving
techniques was analyzed. These results motivate us to find more
FHS sets with optimal AHC and low MHC. Furthermore, it may also be
a challenging problem to find a necessary and sufficient condition
that an FHS set has optimal AHC.

\bibliographystyle{IEEEtran}

\begin{thebibliography}{99}

\bibitem{Blue} Specification of the Bluetooth Systems-Core. The
Bluetooth Special Interest Group (SIG). [Online]. Available:
http://www.bluetooth.com

\bibitem{Beaulieu} N. C. Beaulieu and D. J. Young, ``Designing
time-hopping ultrawidebandwidth receivers for multiuser
interference environments,'' {\em Proc. IEEE}, vol. 97, no. 2, pp.
255-284, 2009.

\bibitem{Vanninen} T. Vanninen, M. Raustia, H. Saarnisaari, J.
Iinatti, ``Frequency hopping mobile ad hoc and sensor network
synchronization,'' {\em Militiary Commun. Conf.}, 2008, pp. 1-7.

\bibitem{Simon} M. K. Simon, J. K. Omura, R. A. Scholtz, and B. K.
Levitt, {\em Spread Spectrum Communications Handbook} (Revised
Ed.). McGraw-Hill Inc., 1994.

\bibitem{Fan} P. Z. Fan and M. Darnell, {\em Sequence Design for
Communications Applications}. Research Studies Press (RSP), John
Wiley $\&$ Sons, London, UK, 1996.

\bibitem{Peng0} D. Peng and P. Fan, ``Lower bounds on the Hamming
auto- and cross correlations of frequency-hopping sequences,''
{\em IEEE Trans. Inform. Theory}, vol. 50, no. 9, pp. 2149-2154,
Sept. 2004.

\bibitem{Lempel} A. Lempel and H. Greenberger, ``Families of
sequences with optimal Hamming correlation properties,'' {\em IEEE
Trans. Inform. Theory}, vol. 20, no. 1, pp. 90-94, Jan. 1974.

\bibitem{Kumar} P. V. Kumar, ``Frequency-hopping code sequence
designs having large linear span,'' {\em IEEE Trans. Inform.
Theory}, vol. 34, no. 1, pp. 146-151, Jan. 1988.

\bibitem{Chu} W. Chu and C. J. Colbourn, ``Optimal
frequency-hopping sequences via cyclotomy,'' {\em IEEE Trans.
Inform. Theory}, vol. 51, no. 3, pp. 1139-1141, Mar. 2005.

\bibitem{Ding} C. Ding and J. Yin, ``Sets of optimal
frequency-hopping sequences,'' {\em IEEE Trans. Inform. Theory},
vol. 54, no. 8, pp. 3741-3745, Aug. 2008.

\bibitem{Han} Y. K. Han  and K. Yang, ``On the Sidel'nikov sequences
as frequency-hopping sequences,'' {\em IEEE Trans. Inform.
Theory}, vol. 55, no. 9, pp. 4279-4285, Sept. 2009.

\bibitem{Chung0} J.-H. Chung, Y. K. Han, and K. Yang, ``New classes of
optimal frequency-hopping sequences by interleaving techniques,''
{\em IEEE Trans. Inform. Theory}, vol. 55, no. 12, pp. 5783-5791,
Dec. 2009.

\enlargethispage{-5.95 in}

\bibitem{Chung11} J.-H. Chung and K. Yang, ``$k$-fold cyclotomy and its
application to frequency-hopping sequences,'' {\em IEEE Trans.
Inform. Theory}, vol. 57, no. 4, pp. 2306-2317, Apr. 2011.

\bibitem{Peng1} D. Peng, X. Niu, and X. Tang, ``Average Hamming
correlation for the cubic polynomial hopping sequences,'' {\em IET
Commun.}, vol. 4, no. 15, pp. 1775-1786, Apr. 2010.

\bibitem{Whiteman} A. L. Whiteman, ``A family of difference
sets,'' {\em Illinois J. Math.}, vol. 6, no. 1, pp. 107-121, Jan.
1962.

\bibitem{Liu} F. Liu, D. Peng, Z. Zhou, and X. Tang, ``Construction of
frequency hopping sequence set based upon generalized cyclotomy,''
arXiv:1009.3602v1.

\bibitem{Golomb} S. W. Golomb and G. Gong, {\em Signal Design for
Good Correlation: for wireless communications, cryptography and
radar applications}. Cambridge University Press, 2005.

\bibitem{Chung1} J.-H. Chung, Y. K. Han, and K. Yang, ``No-hit-zone
frequency-hopping sequence sets with optimal Hamming
autocorrelation,'' {\em IEICE Trans. Fund. Elec. Commun. Comp.
Sci.}, vol. E93-A, no. 11, pp. 2239-2244, Nov. 2010.

\bibitem{Ye1} W. Ye and P. Fan, ``Two classes of frequency
hopping sequences with no-hit zone,'' in {\em Proc. 7th Intl.
Symp. Commun. Theory Appl.}, Ambleside, UK, 2003, pp. 304-306.

\bibitem{Chung2} J.-H. Chung and K. Yang, ``Near-perfect nonlinear
mappings and their applications,'' preprint.

\bibitem{Storer} T. Storer, {\em Cylotomy and Difference Sets,
Lectures in Advanced Mathematics}. Chicago, IL: Markham, 1967.

\bibitem{Sze} T. W. Sze, S. Chanson, C. Ding, T. Helleseth, and M.
G. Parker, ``Logarithm authentication codes,'' {\em Information
and Computation}, vol. 184, no. 1, pp. 93-108, July 2003.

\bibitem{Sidelnikov} V. M. Sidel'nikov, ``Some $k$-valued
pseudo-random sequences and nearly equidistant codes,'' {\em
Probl. Inf. Transm.}, vol. 5, no. 1, pp. 12-16, 1969.


\bibitem{Gong0} G. Gong, ``Theory and applications of $q$-ary
interleaved sequences,'' {\em IEEE Trans. Inform. Theory}, vol.
41, pp. 400-411, Mar. 1995.

\bibitem{Gong1} G. Gong, ``New designs for signal sets with low
cross correlation, balance property, and large linear span:
GF$(p)$ case,'' {\em IEEE Trans. Inform. Theory}, vol. 48, no. 11,
pp. 2847-2867, Nov. 2002.


\bibitem{Zhou} Z. Zhou, X. H. Tang, and G. Gong,
``A new class of sequences with zero or low correlation zone based
on interleaving technique,'' {\em IEEE Trans. Inform. Theory},
vol. 54, no. 9, pp. 4267-4273, Sept. 2008.

\bibitem{Cao} Z. Cao, G. Ge, and Y. Miao, ``Combinatorial
characterizations of one-coincidence frequency-hopping
sequences,'' {\em Des. Codes Crypt.}, vol. 41, no. 2, pp. 177-184,
Nov. 2006.

\end{thebibliography}

\end{document}